\documentclass[graybox]{svmult}

\usepackage{mathptmx}       % selects Times Roman as basic font
\usepackage{helvet}         % selects Helvetica as sans-serif font
\usepackage{courier}        % selects Courier as typewriter font
\usepackage{type1cm}        % activate if the above 3 fonts are
\usepackage{makeidx}         % allows index generation
\usepackage{graphicx}        % standard LaTeX graphics tool
\usepackage{multicol}        % used for the two-column index
\usepackage[bottom]{footmisc}% places footnotes at page bottom
\usepackage{subeqnarray}
\usepackage{mathptmx}
\usepackage{amsfonts}
\usepackage[centertags]{amsmath}
\usepackage{amssymb}
\begin{document}

\title*{Super-de Sitter and alternative super-Poincar\'{e} symmetries
%% Superextension of de Sitter symmetry and alternative super-Poincar\'{e} symmetry
%% $\mathbb{Z}_2\times\mathbb{Z}_2$-graded de Sitter and Poincar\'{e} superalgebras
%% Contribution Title
}
\author{V.N. Tolstoy
%% Name of First Author and Name of Second Author
}
\institute{V.N. Tolstoy \at Lomonosov Moscow State University, Skobeltsyn Institute of
Nuclear Physics (MSU SINP), 1(2) Leninskie Gory, GSP-1,  Moscow 119991, Russian
Federation, \email{tolstoy@nucl-th.sinp.msu.ru}
%% Name of First Author \at Name, Address of Institute, \email{name@email.address} \and Name
%% of Second Author \at Name, Address of Institute \email{name@email.address}
}
\maketitle %%
\abstract{It is well-known that de Sitter Lie algebra $\mathfrak{o}(1,4)$ contrary to
anti-de Sitter one $\mathfrak{o}(2,3)$ does not have a standard $\mathbb{Z}_2$-graded
superextension. We show here that the Lie algebra $\mathfrak{o}(1,4)$ has a
superextension based on the $\mathbb{Z}_2\times\mathbb{Z}_2$-grading. Using the standard
contraction procedure for this superextension we obtain an {\it alternative}
super-Poincar\'{e} algebra with the $\mathbb{Z}_2\times\mathbb{Z}_2$-grading.
%% Each contribution should be preceded by an abstract that summarizes the content.
} %%%%

\section{Introduction}
\label{sec:1}

In supergravity theory (SUGRA) already for more than 20 years there is the following unsolved
(up to now) problem. All physical reasonable solutions of SURGA models with cosmological
constants $\Lambda$ have been constructed for the case $\Lambda<0$, i.e. for the anti-de
Sitter metric
\begin{eqnarray}\label{dS1}
g_{ab}^{}\!\!&=\!\!&\mathop{\rm diag}\,(1,-1,-1,-1,1),\quad (a,b\,=\,0,1,2,3,4)
\end{eqnarray}
with the space-time symmetry $\mathfrak{o}(2,3)$. In the case $\Lambda>0$, i.e. for the
Sitter metric
\begin{eqnarray}\label{dS1} g_{ab}^{}\!\!&=\!\!&\mathop{\rm
diag}\, (1,-1,-1,-1,-1),\quad (a,b\,=\,0,1,2,3,4)
\end{eqnarray}
with the space-time symmetry $\mathfrak{o}(1,4)$ no reasonable solutions have been found.
For example, in SUGRA it was obtained the following relation
\begin{eqnarray}\label{dS1}
\Lambda=-3m^2,
\end{eqnarray}
where $m$ is the massive parameter of gravitinos. Thus if $\Lambda>0$, then $m$ is
imaginary.

In my opinion these problems for the case $\Lambda>0$ are connected with superextensions
of anti-de Sitter $\mathfrak{o}(2,3)$ and de Sitter $\mathfrak{o}(1,4)$ symmetries. The  $\mathfrak{o}(2,3)$ symmetry has the superextension - the superalgebra
$\mathfrak{osp}(1|(2,3))$. This is the usual $\mathbb{Z}_2$-graded superalgebra. In the
case of $\mathfrak{o}(1,4)$ such superextension does not exist. However the Lie algebra
$\mathfrak{o}(1,4)$ has an $alternative$ superextension that is based on the
$\mathbb{Z}_2\times\mathbb{Z}_2$-grading and a preliminary analysis shows that we can
construct the reasonable SUGRA models for the case $\Lambda>0$. In this paper we shall
consider certain $\mathbb{Z}_2\times\mathbb{Z}_2$-graded supersymmetries, but we will not
discuss supergravity models based on such supersymmetries.

All standard relativistic SUSY (super-anti de Sitter, super-Poincar\'{e},
super-confor\-mal, extended $N$-supersymmetry, etc) are based on usual
($\mathbb{Z}_2$-graded) Lie superalgebras ($osp(1|(2,3))$ , $su(N|(2,2))$, $osp(N|(2,3))$
etc). It turns out that every standard relativistic SUSY has an alternative variant based
on an alternative ($\mathbb{Z}_2\times\mathbb{Z}_2$-graded) Lie superalgebra:
\\[5pt]
\makebox{\phantom{aaaaaa}%
\begin{picture}(300,12)\thicklines %
\put(5,0){%\footnotesize
Standard relativistic SUSY}%
\put(153,0){%\footnotesize
Alternative relativistic SUSY}%
\put(120,4){\vector(1,0){25}}%
\put(145,0){\vector(-1,0){25}}%
\end{picture}}

\vspace{0.3cm} Distinctive features of the standard and alternative relativistic
symmetries (in the example of Poincar\'{e} SUSY) are connected with the relations between
the four-momenta and the $Q$-charges and also between the space-time coordinates and the
Grassmann variables. Namely, we have.

%%\noindent
(I) {\it For the standard ($\mathbb{Z}_2$-graded) Poincar\'{e} SUSY:}
\begin{eqnarray}\label{dS1}
[P_{\mu},Q_{\alpha}]&=&[P_{\mu},\bar{Q}_{\dot{\alpha}}]\,=\,0~,\quad
\{Q_{\alpha},\bar{Q}_{\dot{\beta}}\}\;=\;2\sigma_{\alpha\dot{\beta}}^{\mu}P_{\mu},
\\[5pt]
[x_{\mu},\theta_{\alpha}]&=&[x_{\mu},\dot{\theta}_{\dot{\alpha}}]\;=\;
\{\theta_{\alpha},\bar{\theta}_{\dot{\beta}}\}\;=\;0.
\end{eqnarray}

(II) {\it For the altrenative ($\mathbb{Z}_2\times\mathbb{Z}_2$-graded) Poincar\'{e}
SUSY:}
\begin{eqnarray}\label{dS1}
\{P_{\mu},Q_{\alpha}\}&=&\{P_{\mu},\bar{Q}_{\dot{\alpha}}\}\,=\,0,\quad
[Q_{\alpha},\bar{Q}_{\dot{\beta}}]\;=\;2\sigma_{\alpha\dot{\beta}}^{\mu}P_{\mu},
\\[5pt]
\{x_{\mu},\theta_{\alpha}\}&=&\{x_{\mu},\dot{\theta}_{\dot{\alpha}}\}\;=\;
[\theta_{\alpha},\bar{\theta}_{\dot{\beta}}]\;=\;0.
\end{eqnarray}
We wrote down only the relations which are changed in the $\mathbb{Z}_2$- and
$\mathbb{Z}_2\times\mathbb{Z}_2$-cases.

The paper is organized as follows. Section 2 provides definitions and general structure
of $\mathbb{Z}_2$- and $\mathbb{Z}_2\times\mathbb{Z}_2$-graded superalgebras and also
some classification of such simple Lie superalgebras. In Section 3 we describe the
orthosymplectic $\mathbb{Z}_2$- and $\mathbb{Z}_2\times\mathbb{Z}_2$-graded superalgebras
$\mathfrak{osp}(1|4)$ and $\mathfrak{osp}(1|2,2)$ and their real forms. We show here that
a real form of $\mathfrak{osp}(1|4)$  contains $\mathfrak{o}(2,3)$ and a real form of
$\mathfrak{osp}(1|2,2)$ contains $\mathfrak{o}(1,4)$. In Section 4 using the standard
contraction procedure for the superextension $\mathfrak{osp}(1|2,2)$ we obtain an {\it
alternative} super-Poincar\'{e} algebra with the $\mathbb{Z}_2\times
\mathbb{Z}_2$-grading.

\section{$\mathbb{Z}_2$- and $\mathbb{Z}_2\times\mathbb{Z}_2$-graded Lie
superalgebras
%%Preliminaries
} \label{sec:2} {\large\textit{A $\mathbb{Z}_2$-graded superalgebra} \cite{Kac1977}}.
A $\mathbb{Z}_2$-graded Lie superalgebra (LSA) $\mathfrak{g}$, as a linear space, is a
direct sum of two graded components
\begin{eqnarray}\label{sasa1}
\mathfrak{g}&=&\bigoplus_{a=0,1} \mathfrak{g}_{a}\;=
\;\mathfrak{g}_{0}\oplus\mathfrak{g}_{1}
\end{eqnarray}
with a bilinear operation (the general Lie bracket), $[\![\cdot,\cdot]\!]$, satisfying
the identities:
\begin{eqnarray}\label{sasa2}
\deg([\![x_a,y_b]\!])&=&\deg(x_a)+\deg(y_b)\,=\,a+b\quad(\!\!\!\!\!\!\!\mod 2\,), %%
\\[7pt]\label{sasa3}
[\![x_a,y_b]\!]&=&-(-1)^{ab}[\![y_b,x_a]\!], %%
\\[7pt]\label{sasa4}
[\![x_a,[\![y_b,z]\!]]\!]&=&[\![[\![x_a,y_b]\!],z]\!]+
(-1)^{ab}[\![y_b,[\![x_a,z]\!]]\!],
\end{eqnarray}
where the elements $x_a$ and $y_b$ are homogeneous, $x_a\in\mathfrak{g}_{a}$,
$y_b\in\mathfrak{g}_{b}$, and the element $z\in\mathfrak{g}$ is not necessarily
homogeneous. The grading function $\deg(\cdot)$ is defined for homogeneous elements of
the subspaces $\mathfrak{g}_{0}$ and $\mathfrak{g}_{1}$ modulo 2,
$\deg({\mathfrak{g}_{0}})=0$, $\deg({\mathfrak{g}_{1}})=1$. The first identity
(\ref{sasa2}) is called the grading condition, the second identity (\ref{sasa3}) is
called the symmetry property and the condition (\ref{sasa4}) is the Jacobi identity. It  follows from (\ref{sasa2}) that $\mathfrak{g}_{0}$ is a Lie subalgebra in
$\mathfrak{g}$, and $\mathfrak{g}_{1}$ is a $\mathfrak{g}_{0}$-module.  It follows from (\ref{sasa2}) and (\ref{sasa3}) that the general Lie bracket $[\![\cdot,\cdot]\!]$ for
homogeneous elements posses two values: commutator $[\cdot,\cdot]$ and anticommutator
$\{\cdot,\cdot\}$.\\
\noindent 
{\large\textit{A $\mathbb{Z}_2\times\mathbb{Z}_2$-graded superalgebra}
\cite{RittWyl1978a}}. A $\mathbb{Z}_2\times\mathbb{Z}_2$-graded LSA
$\tilde{\mathfrak{g}}$, as a linear space, is a direct sum of four graded components
\begin{eqnarray}\label{sasa5}
\tilde{\mathfrak{g}}\!\!&=\!\!&\bigoplus_{\mathbf{a}=(a_1,a_2)}
\tilde{\mathfrak{g}}_{\mathbf{a}}\;= \;\tilde{\mathfrak{g}}_{(0,0)}\oplus
\tilde{\mathfrak{g}}_{(1,1)}\oplus\tilde{\mathfrak{g}}_{(1,0)}\oplus
\tilde{\mathfrak{g}}_{(0,1)}~
\end{eqnarray}
with a bilinear operation  $[\![\cdot,\cdot]\!]$ satisfying the identities (grading,
symmetry, Jacobi):
\begin{eqnarray}\label{sasa6}
\deg([\![x_{\mathbf{a}},y_{\mathbf{b}}]\!])&=&\deg(x_{\mathbf{a}})+
\deg(y_{\mathbf{b}})\,=\,\mathbf{a}+\mathbf{b}\,=\,(a_1+b_1,a_2+b_2), %%
\\[7pt]\label{sasa7}
[\![x_{\mathbf{a}},y_{\mathbf{b}}]\!]&=&-(-1)^{\mathbf{a}\mathbf{b}}
[\![y_{\mathbf{b}},x_{\mathbf{a}}]\!], %%
\\[7pt]\label{sasa8}
[\![x_{\mathbf{a}},[\![y_{\mathbf{b}},z]\!]]\!]&=&[\![[\![x_{\mathbf{a}},
y_{\mathbf{b}}]\!],z]\!]+(-1)^{\mathbf{a}\mathbf{b}}
[\![y_{\mathbf{b}},[\![x_{\mathbf{a}},z]\!]]\!],
\end{eqnarray}
where the vector $(a_1+b_1,a_2+b_2)$ is defined $\!\!\!\mod\!(2,2)$ and $\mathbf{ab}=
a_1b_1+a_2b_2$. Here in (\ref{sasa6})-(\ref{sasa8}) $x_{\mathbf{a}}\in
\tilde{\mathfrak{g}}_{\mathbf{a}}$, $y_{\mathbf{b}}\in\tilde{\mathfrak{g}}_{\mathbf{b}}$,
and the element $z\in\tilde{\mathfrak{g}}$ is not necessarily homogeneous.  It follows from
(\ref{sasa6}) that $\tilde{\mathfrak{g}}_{(0,0)}$ is a Lie subalgebra in
$\tilde{\mathfrak{g}}$, and the subspaces $\tilde{\mathfrak{g}}_{(1,1)}$,
$\tilde{\mathfrak{g}}_{(1,0)}$ and $\tilde{\mathfrak{g}}_{(0,1)}$ are
$\tilde{\mathfrak{g}}_{(0,0)}$-modules. It should be noted that
$\tilde{\mathfrak{g}}_{(0,0)}\oplus\tilde{\mathfrak{g}}_{(1,1)}$ is a Lie subalgebra in
$\tilde{\mathfrak{g}}$ and the subspace
$\tilde{\mathfrak{g}}_{(1,0)}\oplus\tilde{\mathfrak{g}}_{(0,1)}$ is a
$\tilde{\mathfrak{g}}_{(0,0)}\oplus\tilde{\mathfrak{g}}_{(1,1)}$-module, and moreover
$\{\tilde{\mathfrak{g}}_{(1,1)},\tilde{\mathfrak{g}}_{(1,0)}\}\subset
\tilde{\mathfrak{g}}_{(0,1)}$ and vice versa
$\{\tilde{\mathfrak{g}}_{(1,1)},\tilde{\mathfrak{g}}_{(0,1)}\}\subset
\tilde{\mathfrak{g}}_{(1,0)}$. It follows from (\ref{sasa6}) and (\ref{sasa7})  that
the general Lie bracket $[\![\cdot,\cdot]\!]$ for homogeneous elements posses two values:
commutator $[\cdot,\cdot]$ and anticommutator $\{\cdot,\cdot\}$ as well as in the
previous $\mathbb{Z}_2$-case.

Let us introduce a useful notation of parity of homogeneous elements: {\it the parity
$p(x)$ of a homogeneous element $x$ is a scalar square of its grading $\deg(x)$ modulo
2}. It is evident that for the $\mathbb{Z}_2$-graded superalgebra
$\mathfrak{g}$ the parity coincides with the grading:
$p(\mathfrak{g}_a)=\deg(\mathfrak{g}_a)=\bar{a}$
($\bar{a}=\bar{0},\bar{1}$)\footnote{Integer value of the parity will be denoted with the
bar.}. In the case of the $\mathbb{Z}_2\times\mathbb{Z}_2$-graded superalgebra
$\tilde{\mathfrak{g}}$ we have
\begin{eqnarray}\label{sasa9}
p(\tilde{\mathfrak{g}}_{\mathbf{a}})&:=&\mathbf{a}^2\;=\;a_1^2+a_2^2\qquad
(\!\!\!\!\!\!\!\mod 2\,),
\end{eqnarray}
that is
\begin{eqnarray}\label{sasa10}
p(\tilde{\mathfrak{g}}_{(0,0)})&=&p(\tilde{\mathfrak{g}}_{(1,1)})\;=\;\bar{0}, \qquad
p(\tilde{\mathfrak{g}}_{(1,0)})\;=\;p(\tilde{\mathfrak{g}}_{(0,1)})\;=\;\bar{1}.
\end{eqnarray}
Homogeneous elements with the parity $\bar{0}$ are called even and with parity $\bar{1}$
are odd. Thus,
\begin{eqnarray}\label{sasa11}
\tilde{\mathfrak{g}}\!\!&=\!\!&\tilde{\mathfrak{g}}_{\bar{0}}\oplus
\tilde{\mathfrak{g}}_{\bar{1}},\qquad\tilde{\mathfrak{g}}_{\bar{0}}\,=
\,\tilde{\mathfrak{g}}_{(0,0)}\oplus\tilde{\mathfrak{g}}_{(1,1)},\qquad
\tilde{\mathfrak{g}}_{\bar{1}}\,=\,\tilde{\mathfrak{g}}_{(1,0)}\oplus
\tilde{\mathfrak{g}}_{(0,1)}.
\end{eqnarray}
The even subspace $\tilde{\mathfrak{g}}_{\bar{0}}$ is a subalgebra and the odd one
$\tilde{\mathfrak{g}}_{\bar{1}}$ is a $\tilde{\mathfrak{g}}_{\bar{0}}$-module. Thus the
parity unifies ''cousinly'' the $\mathbb{Z}_2$- and $\mathbb{Z}_2\times
\mathbb{Z}_2$-graded superalgebras. \\%
{\large\textit{Classification of the $\mathbb{Z}_2$- and $\mathbb{Z}_2\times
\mathbb{Z}_2$-graded simple Lie superalgebras}}. A complete list of simple
$\mathbb{Z}_2$-graded (standard) Lie superalgebras was obtained by Kac \cite{Kac1977}.
The following scheme resumes the classification \cite{FraSorSci1996}: %%
\\[7pt]
\makebox
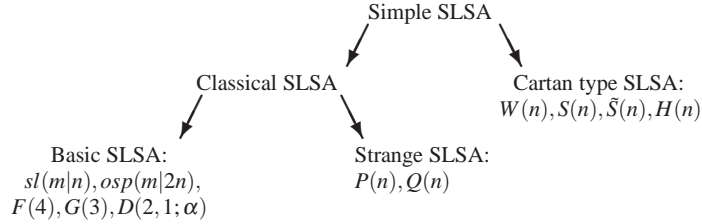
{%%
\begin{picture}(300,10)\thicklines %
\put(150,2){\footnotesize Simple SLSA}%
\put(200,0){\vector(1,-2){8}}%
\put(150,0){\vector(-1,-2){8}}%
\put(205,-25){\footnotesize Cartan type SLSA:}%
\put(200,-35){\footnotesize$W(n),S(n),\tilde{S}(n),H(n)$}%
\put(85,-25){\footnotesize Classical SLSA}%
\put(140,-27){\vector(1,-2){8}}%
\put(87,-27){\vector(-1,-2){8}}%
\put(30,-52){\footnotesize Basic SLSA:}%
\put(20,-62){\footnotesize$sl(m|n),osp(m|2n),$}%
\put(15,-72){\footnotesize$F(4),G(3),D(2,1;\alpha)$}%
\put(145,-52){\footnotesize Strange SLSA:}%
\put(145,-62){\footnotesize$P(n),Q(n)$}%
\end{picture}}%%

\vspace{3.0cm}%%

There is a $\mathbb{Z}_2\times \mathbb{Z}_2$-analog (alternative superalgebras) of this
scheme:
\\[7pt]%
\makebox
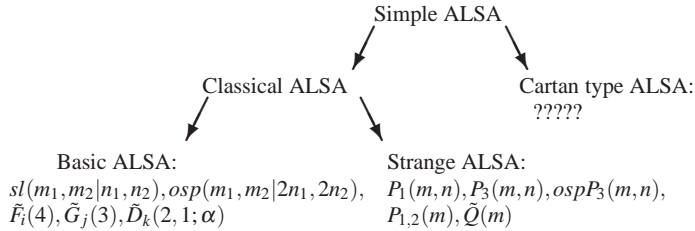
{%%
\begin{picture}(300,10)\thicklines %
\put(150,2){\footnotesize Simple ALSA}%
\put(200,0){\vector(1,-2){8}}%
\put(150,0){\vector(-1,-2){8}}%
\put(205,-25){\footnotesize Cartan type ALSA:}%
\put(210,-35){\footnotesize$?????$} %%W(n),S(n),\tilde{S}(n),H(n)$}%
\put(85,-25){\footnotesize Classical ALSA}%
\put(145,-27){\vector(1,-2){8}}%
\put(87,-27){\vector(-1,-2){8}}%
\put(30,-54){\footnotesize Basic ALSA:}%
\put(12,-64){\footnotesize$sl(m_1,m_2|n_1,n_2),osp(m_1,m_2|2n_1,2n_2),$}%
\put(12,-74){\footnotesize$\tilde{F}_i(4),\tilde{G}_j(3),\tilde{D}_k(2,1;\alpha)$}%
\put(155,-54){\footnotesize Strange ALSA:}%
\put(155,-64){\footnotesize$P_1(m,n),P_3(m,n),ospP_3(m,n),$}%
\put(155,-74){\footnotesize$P_{1,2}(m),\tilde{Q}(m)$}%
\end{picture}}

\vspace{2.9cm}%
\noindent %%
where $i=1,2,\ldots,6$, $j=1,2,3$, $k=1,2,3$. It should be noted that the
classification of the classical series $sl(m_1,m_2|n_1,n_2)$, $osp(m_1,m_2|2n_1,2n_2)$
and all strange series was obtain by Rittenberg and Wyler in \cite{RittWyl1978a}.

There are numerous references about the $\mathbb{Z}_2$-graded Lie superalgebras and their
applications. Unfortunately, in the $\mathbb{Z}_2\times \mathbb{Z}_2$-case the situation
is somewhat poor. There are a few references where some $\mathbb{Z}_2\times
\mathbb{Z}_2$-graded Lie superalgebras were studied and applied
\cite{LukirRitt}--\cite{Zheltukhin}.

Analysis of matrix realizations of the basic $\mathbb{Z}_2\times\mathbb{Z}_2$-graded Lie
superalgebras shows that these superalgebras (as well as the $\mathbb{Z}_2$-graded Lie
superalgebras) have Cartan-Weyl and Chevalley bases, Weyl groups, Dynkin diagrams, etc.
However these structures have a specific characteristics for the $\mathbb{Z}_2$- and
$\mathbb{Z}_2\times\mathbb{Z}_2$-graded cases. Let us consider, for example, the
Dynkin diagrams. In the case of the $\mathbb{Z}_2$-graded superalgebras the nodes of the
Dynkin diagram and corresponding simple roots occur at three types: %%
\\[5pt]
\begin{picture}(100,1)\thicklines %
\put(70,0){white}
\put(100,3){\circle{8}}%
\put(106,0){,}%
\put(125,0){gray}%
\put(150,3){\circle{8}}%
%%\put(115,5){\line(1,-2){5}}%
\put(146.3,0.5){$\times$}
\put(156,0){,}%
\put(175,0){dark}%
\put(200,3){\circle*{8.5}}%
\put(206,0){.}%
\end{picture}
%%}
\\[5pt]
While in the case of $\mathbb{Z}_2\times\mathbb{Z}_2$-graded superalgebras we have six
types of nodes: %%
\\[5pt]%%
\begin{picture}(200,10)\thicklines %
\put(30,0){(00)-white}
\put(80,3){\circle{8}}%
\put(86,0){,}%
\put(100,0){(11)-white}%
\put(150,3){\circle{8}}%
\put(150,7){\line(0,1){3}}%
\put(150,-4.5){\line(0,1){3}}%
%%\put(215,5){\line(1,-2){5}}%
%% \put(245.4,0){$\times$}
\put(156,0){,}%
\put(175,0){(10)-gray}%
\put(220,3){\circle{8}}%
\put(220,7){\line(0,1){3}}%
\put(216.5,0.5){$\times$}
\put(226,0){,}%
%%%%%%%%%%%%
\put(35,-20){(01)-gray}
\put(80,-17){\circle{8}}%
\put(80,-24.5){\line(0,1){3}}%
\put(76.2,-19.5){$\times$}
\put(86,-20){,}%
\put(105,-20){(10)-dark}%
\put(150,-17){\circle*{8.5}}%
\put(150,-13){\line(0,1){3}}%
\put(156,-20){,}%
\put(175,-20){(01)-dark}%
\put(220,-17){\circle*{8.5}}%
\put(220,-24.5){\line(0,1){3}}%
\put(226,-20){.}%
\end{picture}%%
\\[15pt]%%\vspace{2.9cm}

In the next Section we consider in detail two basic superalgebras of rank 2: the
orthosymplectic $\mathbb{Z}_2$-graded superalgebra $\mathfrak{osp}(1|4)$ and the
orthosymplectic $\mathbb{Z}_2\times\mathbb{Z}_2$-graded  superalgebra
$\mathfrak{osp}(1|2,2):=\mathfrak{osp}(1,0|2,2)$. It will be shown that their real forms,
which contain the Lorentz subalgebra $\mathfrak{o}(1,3)$, give us the super-anti-de Sitter
(in the $\mathbb{Z}_2$-graded case) and super-de Sitter (in the $\mathbb{Z}_2\times
\mathbb{Z}_2$-graded case) Lie superalgebras. %%

\section{Anti-de Sitter and de Sitter superalgebras}
\label{sec:3} %%
{\large\textit{The orthosymplectic $\mathbb{Z}_2$-graded superalgebra
$\mathfrak{osp}(1|4)$}}. The Dynkin diagram:
\\[10pt]
\makebox{\phantom{The Dynkin diagram:aaaaaaaa}
$\qquad$
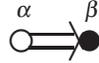
\begin{picture}(1,1)\thicklines %
\put(-10,3){\circle{8}}%
 %% \put(-10,7){\line(0,1){3}}%
 %% \put(-10,-4){\line(0,1){3}}%
\put(-12,12){\footnotesize$\alpha$}%
\put(-7,5.5){\line(1,0){17}}%
\put(-7,0.5){\line(1,0){17}}%
\put(16,3){\circle*{8}}%
\put(8.5,10){\line(1,-2){5}}%
\put(8.5,-4){\line(1,2){5}}%
 %% \put(16,7){\line(0,1){3}}
\put(14,12){\footnotesize$\beta$}%
\end{picture}
}\\[5pt]
The Serre relations:
\begin{eqnarray}\label{dS1}
\begin{array}{rcccl}
[e_{\pm\alpha},[e_{\pm\alpha},e_{\pm\beta}]]&=0&,\qquad
[\{[e_{\pm\alpha},e_{\pm\beta}],e_{\pm\beta}\},e_{\pm\beta}]&=&0.
%%\\[7pt]
\end{array}
\end{eqnarray}
The root system $\Delta_+$:
\begin{eqnarray}\label{dS2}
\begin{array}{rcccl}
\underbrace{2\beta,\;\;2\alpha+2\beta,\;\;\alpha,\;\;\alpha+2\beta}_{\deg(\cdot)=0},\;\;
\underbrace{\beta,\;\;\alpha+\beta}_{\deg(\cdot)=1}.
\end{array}
\end{eqnarray}
\\[5pt]
{\large\textit{The orthosymplectic $\mathbb{Z}_2\times\mathbb{Z}_2$-graded superalgebra
$\mathfrak{osp}(1|2,2)$}}.
%% \\[15pt]
The Dynkin diagram:
\\[10pt]
\makebox{\phantom{The Dynkin diagram:aaaaaaaa}$\qquad$
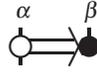
\begin{picture}(1,1)\thicklines %
\put(-10,3){\circle{8}}%
\put(-10,7){\line(0,1){3}}%
\put(-10,-4){\line(0,1){3}}%
\put(-12,14){\footnotesize$\alpha$}%
\put(-7,5.5){\line(1,0){17}}%
\put(-7,0.5){\line(1,0){17}}%
\put(16,3){\circle*{8}}%
\put(8.5,10){\line(1,-2){5}}%
\put(8.5,-4){\line(1,2){5}}%
\put(16,7){\line(0,1){3}}
\put(14,14){\footnotesize$\beta$}%
\end{picture}
} \\[5pt]
The Serre relations:
\begin{eqnarray}\label{dS3}
\begin{array}{rcccl}
\{e_{\pm\alpha},\{e_{\pm\alpha},e_{\pm\beta}\}\}&=&0,\qquad
\{[\{e_{\pm\alpha},e_{\pm\beta}\},e_{\pm\beta}],e_{\pm\beta}\}&=&0. %
\end{array}
\end{eqnarray}%%
The root system $\Delta_+$:
\begin{eqnarray}\label{dS4}
\begin{array}{rcccl}
\underbrace{2\beta,\;\;2\alpha+2\beta}_{\deg(\cdot)=(00)},\;\;
\underbrace{\alpha,\;\;\alpha+2\beta}_{\deg(\cdot)=(11)},\;\;
\underbrace{\beta}_{\deg(\cdot)=(10)},\;\;\underbrace{\alpha+\beta}_{\deg(\cdot)=(01)}.
%% \\[5pt]
\end{array}
\end{eqnarray}
Commutation relations, which contain Cartan elements, are the same for the
$\mathfrak{osp}(1|4)$ and $\mathfrak{osp}(1|2,2)$ superalgebras and they are:
\begin{eqnarray}\label{dS5}
\begin{array}{rcccl}
[\![e_{\gamma},e_{-\gamma'}]\!]&=&\delta_{\gamma,\gamma'}h_{\gamma},
\\[7pt]
[h_{\gamma},\,e_{\gamma'}]&=&(\gamma,\gamma')e_{\gamma'}
\end{array} %%\Bigg\}\quad{\rm for}\;\;\gamma,\gamma'\in\{\alpha,\beta\})
\end{eqnarray}
for $\gamma,\gamma'\in\{\alpha,\beta\}$. These relations together with the Serre
relations (\ref{dS1}) and (\ref{dS3})  correspondingly are called the defining relations of the
superalgebras $\mathfrak{osp}(1|4)$ and $\mathfrak{osp}(1|2,2)$ correspondingly. It is
easy to see that these defining relations  are invariant with respect to the non-graded
Cartan involution $(^{\dag})$ ($(x^{\dag})^{\dag}=x$, $[\![x,y]\!]^{\dag}=
[\![y^{\dag},x^{\dag}]\!]$ for any homogenous elements $x$ and $y$):
\begin{eqnarray}\label{dS6}
\begin{array}{rcccl}
e_{\pm\gamma}^{\dag}&=&e_{\mp\gamma},\qquad h_{\gamma}^{\dag}&=&h_{\gamma}.
\end{array}
\end{eqnarray}
The composite root vectors $e_{\pm\gamma}$ ($\gamma\in\Delta_+$) for
$\mathfrak{osp}(1|4)$ and $\mathfrak{osp}(1|2,2)$ are defined as follows
\begin{eqnarray}\label{dS7}
\begin{array}{rcccl}
e_{\alpha+\beta}&:=&[\![e_{\alpha},\,e_{\beta}]\!],\qquad\qquad\quad
e_{\alpha+2\beta}&:=&[\![e_{\alpha+\beta},\,e_{\beta}]\!],
\\[10pt]
e_{2\alpha+2\beta}&:=&\displaystyle\frac{1}{\sqrt{2}} \{e_{\alpha+\beta},
e_{\alpha+\beta}\},\qquad\;\; e_{2\beta}&:=&\displaystyle
\frac{1}{\sqrt{2}}\{e_{\beta},\,e_{\beta}\},
\\[12pt]
e_{-\gamma}&:=&e_{\gamma}^{\dag}~.\qquad\qquad\qquad\qquad\quad\;\;&&
\end{array}
\end{eqnarray}
These root vectors satisfy the non-vanishing relations:
\begin{eqnarray}\label{dS8}
\begin{array}{rcccl}
[e_{\alpha},e_{\alpha+2\beta}]&=&(-1)^{\deg\alpha\cdot\deg\beta}\sqrt{2}
e_{2\alpha+2\beta},\qquad\;\;\;[e_{\alpha},e_{2\beta}]&=&\sqrt{2}\,e_{\alpha+2\beta},
\\[10pt]
[\![e_{\alpha+\beta},e_{-\alpha}]\!]&=&-(-1)^{\deg\alpha\cdot\deg\beta}e_{\beta},
\qquad\quad\;\;[e_{\alpha+2\beta},e_{-\alpha}]&=&-\sqrt{2}\,e_{2\beta},
\\[10pt]
[e_{2\alpha+2\beta},e_{-\alpha}]&=&-(-1)^{\deg\alpha\cdot\deg\beta}\sqrt{2}
e_{\alpha+2\beta},\qquad[e_{2\beta},e_{-\beta}]&=&-\sqrt{2}\,e_{\beta},
\\[10pt]
[\![e_{\alpha+2\beta},e_{-\alpha-\beta}]\!]&=&-(-1)^{\deg\alpha\cdot\deg\beta}
e_{\beta},\qquad\quad\;\; [\![e_{\beta},e_{-\alpha-\beta}]\!]&=&e_{-\alpha},
\\[10pt]
[\![e_{\beta},e_{-\alpha-2\beta}]\!]&=&-e_{-\alpha-\beta},\qquad\qquad\qquad
[e_{2\alpha+2\beta},e_{-\alpha-\beta}]&=&-\sqrt{2}e_{\alpha+\beta},
\\[10pt]
[e_{\alpha+2\beta},e_{-2\alpha-2\beta}]&=&-(-1)^{\deg\alpha\cdot\deg\beta}
\sqrt{2}e_{-\alpha},\quad\; [e_{2\beta},\,e_{-\alpha-2\beta}]&=&-\sqrt{2}\,e_{-\alpha},
\\[10pt]
\{e_{\alpha+\beta},e_{-\alpha-\beta}\}&=&h_{\alpha}+h_{\beta},\qquad\qquad\qquad\;\;
[e_{\alpha+2\beta},e_{-\alpha-2\beta}]&=&-h_{\alpha}-2h_{\beta},
\\[10pt]
[e_{2\beta},e_{-2\beta}]&=&-2h_{\beta},\qquad\qquad\qquad\quad
[e_{2\alpha+2\beta},e_{-2\alpha-2\beta}]&=&-2h_{\alpha}-2h_{\beta}.
\end{array}
\end{eqnarray}
The rest of non-zero relations is obtained by applying the operation $(^{\dag})$ to these
relations.

Now we find real forms of $\mathfrak{osp}(1|4)$ and $\mathfrak{osp}(1|2,2)$, which
contain the real Lorentz subalgebra $\mathfrak{so}(1,3)$. It is not difficult to check
that the antilinear mapping $(^{*})$ ($(x^*)^*=x$, $[\![x,y]\!]^*=[\![y^*,x^*]\!]$ for any
homogenous elements $x$ and $y$) given by
\begin{equation}\label{dS9}
\begin{array}{rcccl}
e_{\pm\alpha}^*&=&-(-1)^{\deg\alpha\cdot\deg\beta}e_{\mp\alpha}^{},\qquad\quad
e_{\pm\beta}^*&=&-ie_{\pm(\alpha+\beta)}^{},
\\[10pt]
e_{\pm2\beta}^*&=&-e_{\pm(2\alpha+2\beta)}^{},\qquad\qquad\;\;e_{\pm(\alpha+2\beta)}^*&=
&-e_{\pm(\alpha+2\beta)},
\\[10pt]
h_{\alpha}^*&=&h_{\alpha}^{},\qquad\qquad\qquad\qquad\qquad\quad
h_{\beta}^*&=&-h_{\alpha}-h_{\beta}^{}.
\end{array}
\end{equation}
is an antiinvolution and the desired real form with respect to the antiinvolution is
presented as follows. %% ($L_{ab}^*=L_{ab}$ ($a,b = 0,1,2,3,4$)):
%%%%%

\textit{The Lorentz algebra $\mathfrak{o}(1,3)$}:
\begin{eqnarray}\label{dS10}
\begin{array}{rcl}
L_{12}^{}&=&\displaystyle-\frac{1}{2}h_{\alpha}^{},
\\[12pt]
L_{13}^{}&=&\displaystyle-\frac{i}{2\sqrt{2}}\Bigl(e_{2\beta}^{}+
e_{2\alpha+2\beta}+e_{-2\beta}+e_{-2\alpha-2\beta}\Bigr),
\\[12pt] %%\label{dS5}
L_{23}^{}&=&\displaystyle-\frac{1}{2\sqrt{2}}\Bigl(e_{2\beta}^{}-
e_{2\alpha+2\beta}-e_{-2\beta}+e_{-2\alpha-2\beta}\Bigr),%% \qquad
%% L_{12}^{}\;=\;-\frac{1}{2}h_{\alpha}^{},
\\[12pt] %%\label{dS6}
L_{01}^{}&=&\displaystyle\frac{i}{2\sqrt{2}}\Bigl(e_{2\beta}^{}+
e_{2\alpha+2\beta}-e_{-2\beta}-e_{-2\alpha-2\beta}\Bigr), %%\qquad
\\[12pt]
L_{02}^{}&=&\displaystyle\frac{1}{2\sqrt{2}}\Bigl(e_{2\beta}^{}-
e_{2\alpha+2\beta}+e_{-2\beta}-e_{-2\alpha-2\beta}\Bigr),
\\[12pt]
L_{03}^{}&=&\displaystyle-\frac{i}{2}(h_{\alpha}^{}+2h_{\beta}^{}).
\end{array}
\end{eqnarray}

\textit{The generators $L_{\mu4}$}:
\begin{eqnarray}\label{dS11}
\begin{array}{rcl}
L_{04}^{}&=&\displaystyle-\frac{i}{2}\Bigl(e_{\alpha+2\beta}^{}+
(-1)^{\deg\alpha\cdot\deg\beta}e_{-\alpha-2\beta}^{}\Bigr),
\\[12pt]
L_{14}^{}&=&\displaystyle-\frac{i}{2}\Bigl(e_{\alpha}^{}+
(-1)^{\deg\alpha\cdot\deg\beta}e_{-\alpha}\Bigr),
\\[12pt]
L_{24}^{}&=&\displaystyle\frac{1}{2}\Bigl(e_{\alpha}^{}-
(-1)^{\deg\alpha\cdot\deg\beta}e_{-\alpha}\Bigr),
\\[12pt]
L_{34}^{}&=&\displaystyle-\frac{i}{2}\Bigl(e_{\alpha+2\beta}^{}-
(-1)^{\deg\alpha\cdot\deg\beta}e_{-\alpha-2\beta}^{}\Bigr).
\end{array}
\end{eqnarray}
Here are: $\deg\alpha=0,\,\deg\beta=1$, i.e. $(-1)^{\deg\alpha\cdot\deg\beta}=1$, for the
case of the $\mathbb{Z}_2$-grading; $\deg\alpha=(1,1),\,\deg\beta=(1,0)$, i.e.
$(-1)^{\deg\alpha\cdot\deg\beta}=-1$, for the case of the $\mathbb{Z}_2\times
\mathbb{Z}_2$-grading.

The all elements  $L_{ab}$ ($a,b = 0,1,2,3,4$) satisfy the relations
\begin{eqnarray}\label{dS12}
\begin{array}{rcl}
[L_{ab}^{},L_{cd}^{}\bigr]&=&i\bigl(g_{bc}^{}\,L_{ad}^{}-g_{bd}^{}\,L_{ac}^{}+
g_{ad}^{}\,L_{bc}^{}-g_{ac}^{}\,L_{bd}^{}\bigr),
\\[10pt] %%\label{dS11}
L_{ab}^{}&=&-L_{ba}^{},\qquad L^{*}_{ab}\,=\,L_{ab}^{},
\end{array}
\end{eqnarray}
where the metric tensor $g_{ab}^{}$ is given by
\begin{eqnarray}\label{dS13}
\begin{array}{rcl}
g_{ab}^{}&=&\mathop{\rm diag}\,(1,-1,-1,-1,g_{44}^{(\alpha)}),
\\[10pt]\label{dS14}
g_{44}^{(\alpha)}&=&(-1)^{\deg\alpha\cdot\deg\beta}.
\end{array}
\end{eqnarray}
Thus we see that in the case of the $\mathbb{Z}_2$-grading, $(-1)^{\deg\alpha\cdot
\deg\beta}=1$, the generators (\ref{dS10}) and (\ref{dS11}) generate the anti-de-Sitter
algebra $\mathfrak{o}(2,3)$, and in the case of the $\mathbb{Z}_2\times
\mathbb{Z}_2$-grading, $(-1)^{\deg\alpha\cdot\deg\beta}=-1$, the generators (\ref{dS10})
and (\ref{dS11}) generate the de-Sitter algebra $\mathfrak{o}(1,4)$.

Finally we introduce  the "supercharges":
\begin{eqnarray}\label{dS15}
\begin{array}{rcccl}
Q_{1}^{}&:=&\displaystyle\sqrt{2}\,\exp\Bigl(-\frac{i\pi}{4}\Bigr)\;e_{\alpha+\beta}^{},
\qquad Q_{2}^{}&:=&\displaystyle\sqrt{2}\,\exp\Bigl(-\frac{i\pi}{4}\Bigr)\;
e_{-\alpha-\beta}^{},
\\[12pt] %%\label{dS13}
\bar{Q}_{\dot{1}}^{}&:=&\displaystyle\sqrt{2}\,\exp\Bigl(-\frac{i\pi}{4}\Bigr)
\;e_{\beta}^{}~,\quad\qquad
\bar{Q}_{\dot{2}}^{}&:=&\displaystyle\sqrt{2}\,\exp\Bigl(-\frac{i\pi}{4}\Bigr)\; e_{-\beta}^{}.
\end{array}
\end{eqnarray}
They have the following commutation relations between themselves:
\begin{eqnarray}
\begin{array}{rcl}\label{dS16}
\{Q_{1}^{},Q_{1}^{}\}&=&-i2\sqrt{2}e_{2\alpha+2\beta}^{}\,=\,
2(L_{13}-iL_{23}-L_{01}+iL_{02}),
\\[12pt]
\{Q_{2}^{},Q_{2}^{}\}&=&-i2\sqrt{2}e_{-2\alpha-2\beta}^{}\,=\,
2(L_{13}+iL_{23}-L_{01}-iL_{02}),
\\[12pt]
\{Q_{1}^{},Q_{2}^{}\}&=&-i2(h_{\alpha}^{}+h_{\beta}^{})\,=\, 2(L_{03}+iL_{12}),
\\[12pt]
\{\bar{Q}_{\dot{\eta}}^{},\bar{Q}_{\dot{\zeta}}^{}\}&=&\{Q_{\zeta}^{},Q_{\eta}^{}\}^{*}\qquad 
(\bar{Q}_{\dot{\eta}}^{}\,=\,Q_{\eta}^{*}\;\;\rm{for}\;\;\eta=1,2;\;\dot{\eta}=\dot{1},\dot{2}), 
\end{array} %%
\\[12pt]%%
\begin{array}{rcl}\label{dS17}
[\![Q_{1}^{},\bar{Q}_{\dot{1}}^{}]\!]&=&-i2e_{\alpha+2\beta}^{}\,=\,2(L_{04}+L_{34}),
\\[12pt]
[\![Q_{1}^{},\bar{Q}_{\dot{2}}^{}]\!]&=&-i2e_{\alpha}^{}\,=\,2(L_{14}-iL_{24}),
\\[12pt]
[\![Q_{2}^{},\bar{Q}_{\dot{1}}^{}]\!]&=&-i2(-1)^{\deg\alpha\cdot\deg\beta}
e_{-\alpha}^{}\,=\,2(L_{14}+iL_{24}),
\\[12pt]
[\![Q_{2}^{},\,\bar{Q}_{\dot{2}}^{}]\!]&=&-i2(-1)^{\deg\alpha\cdot\deg\beta}
e_{-\alpha-2\beta}^{}\,=\,2(L_{04}-L_{34}). \phantom{aaaaaaa;}
\end{array}
\end{eqnarray}
Here $[\![\cdot,\cdot]\!]\equiv\{\cdot,\cdot\}$ for the $\mathbb{Z}_2$-case and
$[\![\cdot,\cdot]\!]\equiv[\cdot,\cdot]$ for the $\mathbb{Z}_2\times\mathbb{Z}_2$-case.
Using the explicit formulas (\ref{dS10}), (\ref{dS11}), (\ref{dS15}) and the commutation
relations (\ref{dS8}) we can also calculate commutation relations between the
operators $L_{ab}$ and the supercharges $Q$'s and $\bar{Q}$'s .

\section{$\mathbb{Z}_2$- and $\mathbb{Z}_2\times\mathbb{Z}_2$-graded Poincar\'{e}
superalgebras} %%
\label{sec:4} %%%
Using the standard contraction procedure: $L_{\mu4}^{}=R\,P_\mu^{}$ ($\mu=0,1,2,3$),
$Q_{\alpha}^{}\rightarrow\sqrt{R}\;Q_{\alpha}^{}$ and $\bar{Q}_{\dot{\alpha}}^{}\rightarrow\sqrt{R}\;\bar{Q}_{\dot{\alpha}}^{}$ ($\alpha=1,2$; $\dot{\alpha}=\dot{1},\dot{2}$) for $R\rightarrow \infty$ we obtain the super-Poincar\'{e} algebra (standard and alternative) which is generated by $L_{\mu\nu}$, $P_{\mu}$, $Q_{\alpha}$, $\bar{Q}_{\dot{\alpha}}$ where $\mu,\nu=0,1,2,3$; $\alpha=1,2$; $\dot{\alpha}=\dot{1},\dot{2}$, with the relations (we write down only those which are distinguished in the $\mathbb{Z}_2$-
and $\mathbb{Z}_2\times\mathbb{Z}_2$-cases).

(I) {\it For the $\mathbb{Z}_2$-graded Poincar\'{e} SUSY:}
\begin{eqnarray}\label{dS18}
[P_{\mu},Q_{\alpha}]&=&[P_{\mu},\bar{Q}_{\dot{\alpha}}]\,=\,0~,\quad
\{Q_{\alpha},\bar{Q}_{\dot{\beta}}\}\;=\;2\sigma_{\alpha\dot{\beta}}^{\mu}P_{\mu}.
\end{eqnarray}

(II) {\it For the $\mathbb{Z}_2\times\mathbb{Z}_2$-graded Poincar\'{e} SUSY:}
\begin{eqnarray}\label{dS19}
\{P_{\mu},Q_{\alpha}\}&=&\{P_{\mu},\bar{Q}_{\dot{\alpha}}\}\,=\,0~,\quad
[Q_{\alpha},\bar{Q}_{\dot{\beta}}]\;=\;2\sigma_{\alpha\dot{\beta}}^{\mu}P_{\mu},
\end{eqnarray} %

Let us consider the supergroups associated to the $\mathbb{Z}_2$- and
$\mathbb{Z}_2\times\mathbb{Z}_2$-graded Poincar\'{e} superalgebras. A group element $g$
is given by the exponential of the super-Poincar\'{e} generators, namely
\begin{eqnarray}\label{dS120}
g(x^{\mu},\omega^{\mu\nu},\theta^{\alpha},\bar{\theta}^{\dot{\alpha}})\!\!&=\!\!&
\exp(x^{\mu}P_{\mu}+\omega^{\mu\nu}M_{\mu\nu}+\theta^{\alpha}Q_{\alpha}+
\bar{Q}_{\dot{\alpha}}\bar{\theta}^{\dot{\alpha}}).
\end{eqnarray}
Because the grading of the exponent is zero ((0) or (00)) and the result is as follows.

1). {\it $\mathbb{Z}_2$-case}: $\deg P=\deg x =0$, $\deg Q=\deg\bar{Q}=\deg\theta=
\deg\bar{\theta}=1$. This means that
\begin{eqnarray}\label{dS21}
[x_{\mu},\theta_{\alpha}]\!\!&=\!\!&[x_{\mu},\bar{\theta}_{\dot{\alpha}}]\;=\;
\{\theta_{\alpha},\bar{\theta}_{\dot{\beta}}\}\;=\;
\{\theta_{\alpha},\theta_{\beta}\}\;=\;
\{\bar{\theta}_{\dot{\alpha}},\bar{\theta}_{\dot{\beta}}\}\;=\;0.
\end{eqnarray}

2). {\it $\mathbb{Z}_2\times\mathbb{Z}_2$-case}: $\deg P=\deg x=(11)$, $\deg
Q=\deg\theta=(10)$, $\deg\bar{Q}=\deg\bar{\theta}=(01)$. This means that
\begin{eqnarray}\label{dS22}
\{x_{\mu},\theta_{\alpha}\}\!\!&=\!\!&\{x_{\mu},\bar{\theta}_{\dot{\alpha}}\}\;=\;
[\theta_{\alpha},\bar{\theta}_{\dot{\beta}}]\;=\;\{\theta_{\alpha},\theta_{\beta}\}\;=\;
\{\bar{\theta}_{\dot{\alpha}},\bar{\theta}_{\dot{\beta}}\}\;=\;0.
\end{eqnarray}
One defines the superspaces as the coset spaces of the standard and alternative
super-Poincar\'{e} groups by the Lorentz subgroup, parameterized the coordinates
$x^{\mu}$, $\theta^{\alpha}$, $\bar{\theta}^{\dot{\alpha}}$, subject to the condition
$\bar{\theta}^{\dot{\alpha}}= (\theta^{\alpha})^*$. We can define a superfield
$\mathcal{F}$ as a function of superspace.
%%%%%%%%%%%%%%%%%%%
\begin{acknowledgement}
The author would like to thank the Organizers for the kind invitation to speak at the
10-th International Workshop "Lie Theory and Its Applications in Physics" (LT-10, Varna,
June 17-23, 2013), and for support of his visit on the Workshop. The paper was supported
by the RFBR grant No.11-01-00980-a and the grant No.12-09-0064 of the Academic Fund
Program of the National Research University Higher School of Economics. %%
\end{acknowledgement}

\biblstarthook{} %%

\end{document}